\documentclass[11pt]{article}
\usepackage{amsmath, amssymb, amsfonts, amsthm}
\usepackage{epsfig}

\newtheorem{theorem}{Theorem}[section]

\newtheorem{definition}[theorem]{Definition}

\newcommand{\be}{\begin{equation}}
\newcommand{\ee}{\end{equation}}
\newcommand{\bey}{\begin{eqnarray}}
\newcommand{\eey}{\end{eqnarray}}

\newcommand{\E}{{\mathbb E }}
\renewcommand{\P}{{\mathbb P}}

\newcommand{\eps}{\varepsilon}

\newcommand{\bw}{{\bf w}}

\newcommand{\bv}{{\bf v}}
\newcommand{\bu}{{\bf u}}
\newcommand{\bba}{{\bf a}}

\newcommand{\bx}{{\bf x}}
\newcommand{\by}{{\bf y}}

\newcommand{\bR}{{\mathbb R}}
\newcommand{\bC}{{\mathbb C}}
\newcommand{\bN}{{\mathbb N}}

\newcommand{\tr}{\mbox{Tr}}

\newcommand{\wt}{\widetilde}

\newcommand{\const}{\mathrm{const}}

\newcommand{\cN}{{\cal N}}

\input epsf

\newcommand{\donothing}[1]{}

\begin{document}
\title{Spectral Properties of Wigner Matrices}
\author{Benjamin Schlein\\
\\
Institute for Applied Mathematics, University of Bonn\\
Endenicher Allee 60, 53115 Bonn, Germany\\ 
E-mail: benjamin.schlein@hcm.uni-bonn.de}

\maketitle

\begin{abstract}
In these notes we review recent progress (and, in Section \ref{sec:ados}, we announce a new result) concerning the statistical properties of the spectrum of Wigner random matrices.
\end{abstract}

\section{Introduction}
\label{sec:intro}

The general goal of random matrix theory consists in determining the statistical properties of the eigenvalues (and the eigenvectors) of $N \times N$ matrices whose entries are random variables with given laws in the limit $N \to \infty$.

In these notes, we restrict our attention to ensembles of hermitian Wigner matrices, whose entries are, up to the constraints due to hermiticity, independent and identically distributed random variables. This assumption guarantees that all eigenvalues are real. Most of the results that I am going to present can also be extended to ensembles of Wigner matrices with different symmetries (real symmetric and quaternion hermitian ensembles). 

Wigner matrices have been first introduced by Wigner in 1955 to describe the excitation spectra of heavy nuclei. For complex systems like heavy nuclei, it is practically impossible to write down the precise Hamilton operator. For this reason, Wigner assumed the matrix elements of the Hamiltonian to be random variables, and he tried to establish properties of the spectrum holding for almost every realization of the randomness. Remarkably,  Wigner's intuition was confirmed by the experimental data. More generally, the study of the spectrum of Wigner matrices can be considered as a starting point for the analysis of the spectrum of systems with disorder. It is believed, in fact, that the spectrum of Wigner matrices shares many similarity with the spectrum of  more complicated systems, such as random Schr\"odinger operators in the metallic (delocalization) phase. 

\begin{definition}\label{def} An ensemble of hermitian Wigner matrices consists of $N \times N$ matrices $H = (h_{j\ell} )_{1 \leq j,\ell \leq N}$. The entries have the form  
\[ \begin{split} h_{j\ell} &= \frac{1}{\sqrt{N}} (x_{j\ell} + i y_{j\ell}) \qquad \text{for } 1 \leq j < \ell \leq N \\ h_{jj} & = \frac{1}{\sqrt{N}} \, x_{jj} \quad  \quad \hspace{1.2cm}  \text{for } 1 \leq j  \leq N \end{split} \]
where $\{ x_{jj}, x_{j\ell}, y_{j\ell} \}$, $1\leq j \leq N$, $j < \ell \leq N$, is a collection of $N^2$ independent real random variables. We will assume that the variables $\{ x_{j\ell}, y_{j,\ell} \}_{1 \leq j < \ell \leq N}$ have a common law, with $\E \, x_{j\ell} = 0$, and $\E \, x_{j\ell}^2 = 1/2$. We will also assume that the variables $\{ x_{jj} \}_{j=1}^N$ are identically distributed, with $\E \, x_{jj} = 0$, and $\E \, x_{jj}^2 = 1$.
Throughout these notes, we are going to assume sub-Gaussian decay of the entries at infinity, in the sense that there exists $\nu >0$ such that  
\[ \E e^{\nu |x_{j\ell}|^2} < \infty, \qquad \E \, e^{\nu |x_{jj}|^2} < \infty \]
for some (and thus all) $1 \leq j < \ell \leq N$. 
\end{definition}

The assumption of sub-Gaussian decay can be relaxed to subexponential decay, but then some of the statements have to be modified slightly. {F}rom the definition above, we observe that entries of Wigner matrices scale as $N^{-1/2}$ with the dimension $N$. This choice guarantees that, as $N \to \infty$, the spectrum of Wigner matrices remains bounded (in fact, as we will see below, with this choice the spectrum is contained, almost surely in the limit $N \to \infty$, in the interval $[-2,2]$).  
%A simple computation of the expectation of the trace of $H^2$ shows that this choice of %the scaling is indeed correct. On the one hand, we have
%\begin{equation}\label{eq:1} \E \, \tr \, H^2= \E \sum_{i,j} |h_{ij}|^2 = N^2 \E |h_{\ell m}|%^2 \end{equation} for arbitrary $1 \leq \ell \leq m \leq N$. On the other hand \[ \E \tr H^2 %= \E \sum_{\mu=1}^N \mu_\alpha^2 \] where $\mu_1, \dots , \mu_N$ are the $N$ %eigenvalues of $H$. Hence, if we want all eigenvalues to be of order one as $N \to 
%\infty$, $\E \tr H^2$ should be of order $N$. However, from (\ref{eq:1}), this is only %possible if $\E |h_{\ell m }|^2$ is of the order $1/N$. 

Observe that Definition \ref{def} does not specify the precise distribution for the entries of the matrix. In fact, one of the main point of the results that will be presented below is exactly the fact that they apply to a large class of Wigner matrices, and that they are independent of the particular choice for the probability law of the entries. The analysis becomes much easier if one assumes the entries of the matrix to be Gaussian random variables. The Gaussian Unitary Ensemble (GUE) is the ensemble of hermitian Wigner matrices (as defined in Definition \ref{def}) where all entries are Gaussian. In this case, it is possible to write the probability density for the matrix $H$ as
\[ dP (H) = \const \, \cdot e^{-\frac{N}{2} \, \tr H^2} dH \quad \text{where } \qquad  dH = \prod_{1 \leq \ell <m \leq N} d h_{\ell m} d h^*_{\ell m} \prod_{j=1}^N d h_{jj} \] is the product of the Lebesgue measure of the entries of $H$. What makes GUE particularly simple to analyze is its invariance with respect to unitary conjugation. If $H$ is a GUE matrix, and $U$ is an arbitrary (fixed) unitary matrix, then also $UHU^*$ is a GUE matrix. It is possible to show that GUE is the only ensemble of hermitian Wigner matrices which enjoys unitary invariance. Because of the unitary invariance, for GUE it is  possible to find an explicit expression for the probability density function $p_N$ of the $N$ eigenvalues:
\begin{equation}\label{eq:jPDF} p_N  (\mu_1, \dots , \mu_N) = \const \cdot \prod_{i<j}^N (\mu_i - \mu_j)^2 \, e^{-\frac{N}{2} \sum_{j=1}^N \mu_j^2} \, .
\end{equation}
The Gaussian factor reflects the Gaussian distribution of the entries. The correlations among the eigenvalues are described by the Vandermond determinant squared  
\begin{equation}\label{eq:VDM}  \prod_{i<j}^N (\mu_i - \mu_j)^2 = \Delta (\mu_1, \dots ,\mu_N)^2 = \left[  \det \left(\begin{array}{lll} 1 & \dots & 1 \\ \mu_1 & \dots & \mu_N \\ & \dots & \\ 
\mu_1^{N-1} & \dots & \mu_N^{N-1} \end{array} \right) \right]^2 \, . \end{equation}
%It is clear from this expression that the correlations among the eigenvalues is repulsive; %if the distance between two neighboring eigenvalues tends to zero, the square of the %Vandermonde determinant converges to zero quadratically. 
Using this explicit expression for the joint probability density function of the $N$ eigenvalues of GUE matrices, one can compute the local eigenvalue statistics. 
For fixed $k=1,\dots , N$, we define the $k$ point correlation function 
\[ p_N^{(k)} (\mu_1 , \dots , \mu_k) = \int d\mu_{k+1} \dots d\mu_N \, p_N (\mu_1, \dots \mu_N) \, . \]
It turns out that, as $N \to \infty$, the local correlation functions of GUE converge, after appropriate rescaling, to the Wigner-Dyson distribution. For any fixed $k \in \bN$,  
\begin{equation}\label{eq:WD} \frac{1}{\rho_{\text{sc}}^k (E)} \, p_N^{(k)} \left( E+ \frac{x_1}{N\rho_{\text{sc}} (E)}, \dots , E + \frac{x_k}{N\rho_{\text{sc}} (E)} \right) \to \det \left( \frac{\sin (\pi (x_i - x_j))}{(\pi (x_i -x_j))} \right)_{1 \leq i,j \leq k} \,  \end{equation}
as $N \to \infty$. Here \begin{equation}\label{eq:sc} \rho_{\text{sc}} (E) = \left\{ \begin{array}{ll} \frac{1}{2\pi} \sqrt{1- \frac{E^2}{4}} \qquad &\qquad \text{if } |E| \leq 2 \\ 0 \qquad & \qquad \text{otherwise} \end{array} \right. \end{equation} is the {\it semicircle law}, which describes the density of the eigenvalues around $E$ (we will discuss the semicircle law in more details in Section \ref{sec:sc}). The convergence in (\ref{eq:WD}) is pointwise in $E$ and holds weakly in the variables $x_1, \dots , x_k$. One refers to the left hand side of (\ref{eq:WD}) as {\it local} correlations because the arguments of $p^{(k)}_N$ vary in an interval of size $1/N$. As we discussed above, the choice of the $N$ dependence of the matrix entries implies that all eigenvalues are contained in a finite interval; this means that the typical distance between neighboring eigenvalues is of the order $N^{-1}$. Hence, to observe non trivial correlations, the relevant length scale is exactly of the order $N^{-1}$. Eq. (\ref{eq:WD}) was first proven for GUE by Dyson in \cite{D} using the explicit expression (\ref{eq:jPDF}) and the asymptotics of Hermite polynomials. Although (\ref{eq:jPDF}) only holds true for GUE, and there is no explicit expression for the joint probability density function of the eigenvalues of any other ensembles of Wigner matrices, the convergence (\ref{eq:WD}) is expected to hold true independently of the law of the matrix entries. One expects, in other words, the local eigenvalue correlations to be {\it universal}; we will present a proof of universality in Section \ref{sec:univ}. In fact, universality should hold even more generally; the Wigner-Dyson statistics is expected to describe the local eigenvalue correlations in a large class of systems with disorder (for example, random Schr\"odinger operators in the metallic phase). 

\section{Density of States and the Semicircle Law}
\label{sec:sc}

The first rigorous result in random matrix theory has been obtained in \cite{W}, where Wigner proved the convergence of the density of states to the semicircle law for arbitrary ensembles of Wigner matrices (hermitian, real symmetric, or quaternion hermitian). For $a < b$, let $\cN [a;b]$ denote the number of eigenvalues in the interval $[a;b]$. The density of the eigenvalues, or density of states, in the interval $[a;b]$ is defined as \[ \rho_{[a;b]} = \frac{\cN [a;b]}{N|b-a|}\, ,  \] where the factor of $N$ in the denominator makes sure that, typically, $\rho_{[a;b]}$ is a quantity of order one. The density of states $\rho_{[a;b]}$ is a random variable whose precise value depend on the realization of the randomness. Nevertheless, Wigner proved that, as $N \to \infty$, $\rho_{[a;b]}$ approaches a deterministic limit. More precisely, he showed that, for any $\delta >0$, 
\[ \lim_{N \to\infty} \P \left( \left| \frac{\cN [a;b]}{N|b-a|} - \frac{1}{|b-a|} \int_a^b ds \, \rho_{\text{sc}} (s) \right| \geq \delta \right) = 0 \]
where $\rho_{\text{sc}} (s)$ denotes the semicircle law defined in (\ref{eq:sc}).
After considering the limit $N \to \infty$, we can also let $\eta = |b-a| \to 0$; in this limit the density of state converges to the semicircle law at $E = (a+b)/2$. Hence, Wigner's result can be formulated as 
\[ \lim_{\eta \to 0}  \lim_{N \to\infty} \P \left( \left| \frac{\cN \left[E-\frac{\eta}{2};E+\frac{\eta}{2}\right]}{N\eta} - \rho_{\text{sc}} (E) \right| \geq \delta \right) = 0 \, .\]
It is worth noticing that the semicircle law, and therefore the limiting density of states, is independent of the choice of the probability law for the entries of the Wigner matrix. It is also important to observe that Wigner's result concerns the density of states on {\it macroscopic} intervals, that is intervals which typically contain order $N$ eigenvalues (for fixed $\eta >0$, the interval $[E-(\eta/2); E+(\eta/2)]$ contains order $N$ eigenvalues).   What happens if we consider the density of states in smaller intervals, namely in intervals whose size shrinks to zero as $N \to \infty$? These intervals will not contain order $N$ eigenvalues. However, as long as the number of eigenvalues is large as $N \to \infty$, we may expect the fluctuations of the density of states to be negligible in the limit. This is the content of the next theorem, which was proven in \cite{ESY3} (extending previous results from \cite{ESY1,ESY2}). 
\begin{theorem}\label{thm:sc}
Consider an ensemble of hermitian Wigner matrices as in Def. \ref{def}. Let $|E|<2$. Then, for any $K >0$, 
\[ \P \left( \left| \frac{\cN \left[E-\frac{K}{2N} ; E+\frac{K}{2N} \right]}{K} - \rho_{\text{sc}} (E) \right| \geq \delta \right) \lesssim e^{-c\delta^2 \sqrt{K}} \] uniformly in $N$, for all $N$ large enough. This quantitative bound implies, in particular, that
\begin{equation}\label{eq:microsc} \lim_{K \to \infty} \lim_{N \to \infty} \P \left( \left| \frac{\cN \left[E-\frac{K}{2N} ; E+\frac{K}{2N} \right]}{K} - \rho_{\text{sc}} (E) \right| \geq \delta \right) = 0 \, . \end{equation}
\end{theorem}
This theorem establishes convergence of the density of states to the semicircle law on {\it microscopic} intervals, that is on intervals containing, typically, a constant ($N$ independent) number of eigenvalues.  {F}rom convergence on the microscopic scale, we also obtain convergence to the semicircle law on intermediate scales; for any sequence $\eta (N) >0$ such that $\eta (N) \to 0$ and $N \eta (N) \to \infty$ as $N \to \infty$ we have
\[ \lim_{N \to \infty} \P \left( \left| \frac{\cN \left[E-\frac{\eta(N)}{2} ; E+\frac{\eta (N)}{2} \right]}{N\eta (N)} - \rho_{\text{sc}} (E) \right| \geq \delta \right) = 0\, . \]
Note that, if $\eta (N) \lesssim 1/N$, the fluctuations of the density of states are certainly important, and one cannot expect convergence in probability; in this sense, (\ref{eq:microsc}) establishes convergence to the semicircle law on the optimal scale.

The proof of Theorem \ref{thm:sc} relies on two main ingredients, an upper bound on the density of states and a fixed point equation for the Stieltjes transform of the empirical eigenvalue distribution. 

\medskip

{\it Upper bound.} Consider a sequence $\eta (N)\geq (\log N)^2/N$. Then there are constants $K_0,C,c$ such that  
\begin{equation}\label{eq:upper} \P \left( \frac{\cN \left[E- \frac{\eta (N)}{2} ; E + \frac{\eta (N)}{2} \right]}{N \eta (N)} \geq K \right) \lesssim e^{-c\sqrt{K N \eta (N)}} \end{equation}
for all $K>K_0$ and $N$ large enough. 

A similar statement is valid also for smaller intervals, of size $2/N \leq \eta (N) \leq (\log N)^2/N$, but its proof is more involved (see Theorem 5.1 in \cite{ESY3}).

To show (\ref{eq:upper}), we observe that (using $\eta \equiv \eta (N)$) 
\[\begin{split} \cN \left[ E-\frac{\eta}{2} ; E + \frac{\eta}{2} \right] &= \sum_{\alpha=1}^N {\bf 1} (|\mu_\alpha - E| \leq \eta/2) \lesssim \eta \, \text{Im} \sum_{\alpha=1}^N \frac{1}{\mu_\alpha - E - i \eta} \end{split} \]
where $\{ \mu_\alpha \}_{\alpha=1}^N$ are the eigenvalues of $H$. Hence \begin{equation}\label{eq:DOS-res} \frac{\cN \left[ E-\frac{\eta}{2} ; E + \frac{\eta }{2} \right] }{N\eta} \lesssim \frac{1}{N} \text{Im } \tr \, \frac{1}{H-E-i\eta}  = \frac{1}{N} \text{Im } \sum_{j=1}^N \frac{1}{H-E - i\eta} (j,j) \,. \end{equation}
To bound the diagonal entries of the resolvent, we use (for the case $j=1$):
\begin{equation}\label{eq:schur1} \frac{1}{H-E-i\eta} (1,1) = \frac{1}{h_{11} - E - i \eta - \langle \bba, (B-E-i\eta)^{-1} \bba \rangle} \end{equation}
where $\bba = (h_{21}, h_{31}, \dots , h_{N1}) \in \bC^{N-1}$ is the first row of $H$ after removing the $(1,1)$-entry, and $B$ is the $(N-1) \times (N-1)$ minor of $H$ obtained by removing the first row and the first column. Using the spectral decomposition of the minor $B$, we find, 
\begin{equation}\label{eq:schur2} \frac{1}{H-E-i\eta} (1,1) = \frac{1}{h_{11} - E - i \eta - \frac{1}{N} \sum_{\alpha=1}^{N-1} \frac{\xi_\alpha}{\lambda_\alpha -E-i\eta}} \end{equation}
where $\xi_\alpha = N | \bba \cdot \bu_\alpha|^2$ and where $\{ \lambda_\alpha \}_{\alpha=1}^{N-1}$ and $\{ \bu_\alpha \}_{\alpha=1}^{N-1}$ are the eigenvalues and, respectively, the eigenvectors of the minor $B$. We conclude that \begin{equation}\label{eq:schur-Im} \text{Im } \frac{1}{H-E-i\eta} (1,1)  \leq \frac{1}{\eta + \text{Im} \frac{1}{N} \sum_{\alpha=1}^N \frac{\xi_\alpha}{\lambda_\alpha -E -i \eta}}  \leq \frac{1}{\text{Im} \frac{1}{N} \sum_{\alpha=1}^N \frac{\xi_\alpha}{\lambda_\alpha -E -i \eta}}\, . \end{equation}
Since $\bba$ is independent of the minor $B$ (and therefore of its eigenvectors), we have \[ \E \, \xi_\alpha = N \E \sum_{i,j} a_i \overline{a}_j \overline{u}_\alpha (j) u_\alpha (i) = N \E \sum_{i,j} \frac{\delta_{ij}}{N} \,  \overline{u}_\alpha (j) u_\alpha (i) = 1 \, . \]
Moreover, it turns out that $\xi_\alpha$ is well concentrated around its expectation. 
Therefore, up to a set with small probability, we find from (\ref{eq:schur-Im}) that 
\begin{equation}\label{eq:schur-Im2} \begin{split} \text{Im } \frac{1}{H-E-i\eta} (1,1)  & \lesssim \frac{K^2}{\text{Im} \frac{1}{N} \sum_{\alpha=1}^N \frac{1}{\lambda_\alpha -E -i \eta}} = \frac{K^2}{\text{Im} \frac{1}{N} \tr \frac{1}{B-E-i\eta}}
\end{split}\end{equation}
for a sufficiently large constant $K >0$. A more precise analysis shows that the measure of the excluded set (on which the bound (\ref{eq:schur-Im2}) may fail) is at most $\exp (-c \sqrt{K N\eta})$; see Lemma 4.7 in \cite{ESY3}. Using (\ref{eq:DOS-res}), but for $B$ instead of $H$, we obtain 
\[ \begin{split} \text{Im } \frac{1}{H-E-i\eta} (1,1)  & \lesssim \frac{K^2}{\frac{\cN_B \left[ E-\frac{\eta}{2}; E+\frac{\eta}{2} \right]}{ N \eta}} \lesssim \frac{K^2}{\frac{\cN \left[ E-\frac{\eta}{2}; E+\frac{\eta}{2} \right]}{ N \eta}}
\end{split}\]
where $\cN_B [a;b]$ denotes the number of eigenvalues of $B$ in the interval $[a;b]$ and where, in the second inequality, we used the fact that the eigenvalues of $B$ are interlaced between the eigenvalues of $H$. 

Analogously, for any $j=2,\dots ,N$, we find that 
\begin{equation}\label{eq:schurj}  \text{Im } \frac{1}{H-E-i\eta} (j,j)  \lesssim \frac{K^2}{\frac{\cN \left[ E-\frac{\eta}{2} ; E + \frac{\eta}{2} \right] }{N\eta}} \end{equation}
up to a set with probability smaller than $\exp (-c \sqrt{KN\eta})$. Since, by assumption,  $N \eta \geq (\log N)^2$, the total measure of the union of these $N$ ``bad'' sets is bounded by $\exp (-(c/2) \sqrt{KN \eta (N)})$. On the complementary set, (\ref{eq:schurj}) is correct for all $j =1,\dots ,N$, and thus, from (\ref{eq:DOS-res}), we find 
\[ \frac{\cN \left[ E-\frac{\eta}{2} ; E + \frac{\eta}{2} \right] }{N\eta} \lesssim \frac{K^2}{\frac{\cN \left[ E-\frac{\eta}{2} ; E + \frac{\eta}{2} \right] }{N\eta}} \qquad \Rightarrow  \qquad \frac{\cN \left[ E-\frac{\eta}{2} ; E + \frac{\eta}{2} \right] }{N\eta} \lesssim K \]
which shows the upper bound (\ref{eq:upper}). 

{\it Stieltjes transform.} We define the Stieltjes transform of the empirical eigenvalue measure by
\[ m_N (z) = \frac{1}{N} \tr \, \frac{1}{H-z} = \frac{1}{N} \sum_{\alpha=1}^N \frac{1}{\mu_\alpha -z} \, .\]
We are going to compare $m_N (z)$ with the Stieltjes transform of the semicircle law, given by
\[ m_{sc} (z) = \int \frac{ds \rho_{sc} (s)}{s-z} = -\frac{z}{2} + \sqrt{\frac{z^2}{4} -1}\,. \]
It turns out that Theorem \ref{thm:sc} follows, if we can show that the difference $m_N (z)-m_{sc} (z)$ converges to zero as $N \to \infty$, for all $z = E+i\eta$ with $|E| <2$ and $\eta \geq K/N$ (see the proof of Corollary 4.2 in \cite{ESY1}). 
%In the proof of the upper bound (\ref{eq:upper}), we already used the fact that the %density of states in the interval of size $\eta$ around $E$ can be bounded above by the %imaginary part of $m_N (E+i\eta)$. It turns out, actually, that the density of states on an %interval of size $\eta$ can be reconstructed from the Stieltjes transform $m (E+i \delta %\eta)$,  with an error which vanishes as $\delta \to 0$ (see the proof of Corollary 4.2 in %\cite{ESY1}). Hence, Theorem \ref{thm:sc} follows if we can show convergence of the %Stieltjes transform $m_N (z)$ to $m_{sc} (z)$ for all $z= E+i\eta (N)$, with $|E| < 2$ %and $\eta (N) \geq K/N$. 
It is worth noticing that $m_{sc} (z)$ satisfies the fixed point equation 
\begin{equation}\label{eq:fixed} m_{sc} (z) + \frac{1}{z+m_{sc} (z)} = 0 \,. \end{equation}
It turns out that this equation is stable away from the spectral edges $E= \pm 2$. Hence to show that $|m_N (z) - m_{sc} (z)| \leq C \delta$ with high probability, it is enough to prove that $m_N (z)$ is an approximate solution of (\ref{eq:fixed}), that is that 
\begin{equation}\label{eq:approx-sol} \left| m_N (z) + \frac{1}{z+m_N (z)} \right| \leq \delta \end{equation} 
with high probability. 

To establish (\ref{eq:approx-sol}), we use again the expression (\ref{eq:schur2}):
\begin{equation}\label{eq:mN-schur} \begin{split}
m_N (z) = \frac{1}{N} \sum_{j=1}^N \frac{1}{H-z} (j,j) = \frac{1}{N} \sum_{j=1}^N \frac{1}{h_{jj} - z - \frac{1}{N} \sum_{\alpha=1}^{N-1} \frac{\xi^{(j)}_\alpha}{\lambda^{(j)}_\alpha - z}} 
\end{split}
\end{equation}
where $\xi^{(j)} = N |\bba^{(j)} \cdot \bu^{(j)}_\alpha|^2$, $\bba^{(j)} = (h_{j1}, \dots, h_{j,j-1}, h_{j,j+1}, h_{jN}) \in \bC^{N-1}$ is the $j$-th row of $H$ without diagonal element, and where $\lambda^{(j)}_\alpha$ and $\bu^{(j)}_\alpha$ are the eigenvalues and the eigenvectors of the minor $B^{(j)}$ obtained from $H$ by removing the $j$-th row and the $j$-th column. {F}rom (\ref{eq:mN-schur}), we obtain 
\begin{equation}\label{eq:mN-schur2} \begin{split} m_N (z) &= \frac{1}{N} \sum_{j=1}^N \frac{1}{- z - m_N (z) -X^{(j)} (z)} \\
&= \frac{-1}{z+ m_N(z)} + \frac{1}{N} \sum_{j=1}^N \frac{X^{(j)}(z)}{(z+m_N (z)) (z + m_N (z) + X^{(j)} (z))} \, .
\end{split}
\end{equation}
Here we defined
\[ X^{(j)} (z) = -h_{jj}  + \left(\frac{N-1}{N} m^{(j)}_{N-1} (z) - m_N (z) \right) + \frac{1}{N} \sum_{\alpha=1}^{N-1} \frac{\xi_\alpha-1}{\lambda_\alpha-z} \]
where 
\[ m_N^{(j)} (z) = \frac{1}{N-1} \tr \, \frac{1}{B^{(j)} - z} = \frac{1}{N-1} \sum_{\alpha=1}^{N-1} \frac{1}{\lambda^{(j)}_\alpha - z} \] is the Stieltjes transform of the minor $B^{(j)}$.

{F}rom (\ref{eq:mN-schur2}) we obtain (\ref{eq:approx-sol}) if we can show that $X^{(j)} (z)$ is small with high probability. The first summand in $X^{(j)} (z)$, $-h_{jj}$, is of the order $N^{-1/2}$ and therefore small. The second summand in $X^{(j)} (z)$,  the difference $(\frac{N-1}{N} m^{(j)}_{N-1} (z) - m	_N (z))$, is small (with probability one); this is a consequence of the interlacing between the eigenvalues of $B^{(j)}$ and $H$.  The main difficulty consists therefore in showing the smallness of the random variable
\begin{equation}\label{eq:Y} Y^{(j)} (z) = \frac{1}{N} \sum_{\alpha=1}^{N-1} \frac{\xi_\alpha - 1}{\lambda_\alpha -z} \, .\end{equation}
To show that $Y^{(j)} (z)$ is small, with high probability, we use the upper bound (\ref{eq:upper}), which guarantees that the denominator in (\ref{eq:Y}) can only be large for a small number of $\alpha$'s; details can be found in Section 6 of \cite{ESY3}.

\section{Delocalization of Eigenvectors of Wigner Matrices}
\label{sec:deloc}

As a first application of Theorem \ref{thm:sc}, we show delocalization of the eigenvectors of Wigner matrices. Given a vector $\bv \in \bC^N$ with $\ell^2$ norm equal to one, we say that it is completely {\it localized} if one of its component has size one, and all other  vanish. $\bv$ is called completely {\it delocalized}, if all its components have the same size (namely $N^{-1/2}$). To distinguish between localized and delocalized vectors, we can compute the $\ell^p$ norm of $\bv$, for $p >2$. If $\bv$ is completely localized, its $\ell^p$ norm equals to one, for all $2 < p \leq \infty$, and for all $N$. On the other hand, if $\bv$ is completely delocalized, 
\[ \| \bv \|_p = N^{-\frac{1}{2} + \frac{1}{p}} \] and therefore converges to zero, as $N \to \infty$ (for all $p>2$). The next theorem was proved in \cite{ESY3}, extending results from \cite{ESY1,ESY2}. 
\begin{theorem}\label{thm:deloc}
Consider an ensemble of hermitian Wigner matrices as in Def. \ref{def}. Let $|E|<2$, $K >0$, $2 < p < \infty$. Then there exists $c,C>0$ such that
\begin{equation}\label{eq:deloc1} \P \left( \exists \bv : H \bv = \mu \bv, |\mu - E| \leq \frac{K}{2N}, \| \bv \|_2 = 1, \|\bv \|_p \geq M N^{-\frac{1}{2}+ \frac{1}{p}} \right) \leq C e^{-c\sqrt{M}} \end{equation}
for all $M > 0$, and $N$ large enough. 
\end{theorem}
If we take $M > (\log N)^2$, the statement can be modified as follows. For every $\kappa >0$, and $2 < p \leq \infty$, there exist $c,C >0$ such that 
\begin{equation}\label{eq:deloc2} \P \left( \exists \bv : H \bv = \mu \bv, |\mu| \leq 2-\kappa, \| \bv \|_2 = 1, \|\bv \|_p \geq M N^{-\frac{1}{2}+ \frac{1}{p}} \right) \leq C e^{-c\sqrt{M}} \end{equation} for all $M >(\log N)^2$, and all $N$ large enough. 

The interpretation of (\ref{eq:deloc1}) and (\ref{eq:deloc2}) is straightforward; eigenvectors of Wigner matrices are completely delocalized. Up to the parameter $M>0$, which tunes the probability, all components of the eigenvectors have the same size.  

The proof of (\ref{eq:deloc2}) is a simple application of Theorem \ref{thm:sc} ((\ref{eq:deloc1}) requires a little bit more work, but the main ideas are the same). For $p=\infty$, we need an upper bound on the components of an eigenvector $\bv = (v_1, v_2, \dots , v_N)$. To bound, for example, $v_1$, we write $\bv = (v_1, \bw)$, with $\bw = (v_2, \dots , v_N)$. {F}rom $H\bv = \mu \bv$, we find 
\[ \bw = - v_1 (B-\mu)^{-1} \bba \]
where $\bba = (h_{21}, \dots , h_{N1})$ is the first column of $H$, after removing the first component, and where $B$ is the $(N-1) \times (N-1)$ minor of $H$ obtained by removing the first row and column. The normalization condition $v_1^2 + \bw^2 = 1$ implies that 
\[ \begin{split} v_1^2  & = \frac{1}{1+ \langle \bba, (B-\mu)^{-2} \bba \rangle} = \frac{1}{1+ \frac{1}{N} \sum_{\alpha =1}^N \frac{\xi_\alpha}{(\lambda_\alpha - \mu)^2}} \leq \frac{1}{\frac{1}{N} \sum_{\alpha =1}^N \frac{\xi_\alpha}{(\lambda_\alpha - \mu)^2}} \end{split} \]
where, as in Section \ref{sec:sc}, $\xi_\alpha = N |\bu_\alpha \cdot \bba|^2$, and $\lambda_\alpha$ and $\bu_\alpha$ are the eigenvalues and the eigenvectors of the minor $B$. Restricting the sum to those $\alpha$ for which $|\lambda_\alpha - \mu| \leq K/N$, and using the fact that $\xi_\alpha$ concentrates around one, we obtain that, with high probability, 
\[   v_1^2  \lesssim \frac{K^2}{N} \frac{1}{\left| \{ \alpha : |\lambda_\alpha - \mu| \leq K/N \} \right|}  \lesssim \frac{K}{N} \]
where, in the last inequality, we use the fact that, by Theorem \ref{thm:sc}, the density of states on the interval $[\mu-(K/N) ; \mu + (K/N)]$ converges, with high probability, to $\rho_{sc} (\mu) >0$. This shows (\ref{eq:deloc2}).

\section{Bulk Universality for Hermitian Wigner Matrices}
\label{sec:univ}

Universality for hermitian Wigner matrices refers to the fact that the local eigenvalue statistics are independent of the particular choice for the probability law of the entries of the matrix. One should distinguish between local correlations close to the edges and in the bulk of the spectrum. The correlations at the edges are described by the Tracy-Widom distribution; for GUE this was first observed in \cite{TW}. In \cite{S}, Soshnikov established {\it edge universality} of the Tracy-Widom distribution (recently, Tao and Vu found a new proof of edge universality in \cite{TV2}). In these notes, I will restrict my attention to universality in the bulk of the spectrum; to show {\it bulk universality}, we need to prove the convergence of the local eigenvalues correlations to the Wigner-Dyson distribution (\ref{eq:WD}).  

A first partial result towards bulk universality for hermitian Wigner matrices was obtained by Johansson, who proved, in \cite{J}, convergence to the Wigner-Dyson statistics for ensembles of Wigner matrices with a Guassian component. Johansson considered Wigner matrices of the form
\begin{equation}\label{eq:Joh} H = H_0 + t^{1/2} V \end{equation} 
where $H_0$ is an arbitrary Wigner matrix, $V$ is a GUE matrix, independent of $H_0$, and $t>0$ measures the strength of the Gaussian component. Strictly speaking, in order for $H$ to obey Definition \ref{def}, we should renormalize it by a $t$ dependent factor to make sure that the variance of the entries remains constant. However, this rescaling only affects the eigenvalue statistics in a trivial way, and therefore in the following we will work with the matrix $H$, as defined in (\ref{eq:Joh}). 

It turns out that $H$ emerges from $H_0$ by letting every entry evolve through independent Brownian motions up to time $t$. This stochastic evolution of the matrix entries induces a stochastic dynamics for the eigenvalues, known as the {\it Dyson Brownian motion}. The Dyson Brownian motion can be described as a flow for the density of the eigenvalues. 
%, which can be described by the system of $N$ coupled stochastic differential equations
%\[ d\mu_\alpha = \frac{dB_\alpha}{\sqrt{N}} + \sum_{\beta \not = \alpha} \frac{1}{\mu_
%\alpha - \mu_\beta} dt \]
%for $\alpha = 1, \dots , N$. Here $\{ B_\alpha \}_{\alpha=1}^N$ is a collection of $N$ %independent Brownian motions. Dyson Brownian motion can also be described as a %flow for the probability density of the eigenvalues. In fact, 
The joint probability density function of the $N$ eigenvalues of $H=H_0 + t^{1/2} V$ can be written as
\begin{equation}\label{eq:jointH} p_{N,t} (\bx) = \int d\by \, q_t (\bx ; \by) p_{N,0} (\by)  \end{equation}
where $\bx = (x_1, \dots ,x_N)$, $\by = (y_1, \dots, y_N)$, $p_{N,0}$ is the joint probability density function of the eigenvalues of the initial matrix $H_0$, and $q_t$ is the kernel describing the Dyson Brownian Motion (describing, in other words, the addition of a GUE matrix to $H_0$). The kernel $q_t$ can be computed explicitly:
\begin{equation}\label{eq:qt-her} q_t (\bx ; \by) = \frac{N^{N/2}}{(2\pi t)^{N/2}} \; \frac{\Delta (\bx)}{\Delta (\by)}
\;  \det \big( e^{-N(x_j-y_k)^2/ 2t}\big)_{j,k=1}^N, \end{equation}
with the Vandermonde determinant $\Delta (\bx)$ defined in (\ref{eq:VDM}). 
{F}rom (\ref{eq:jointH}), we can express the $k$-point correlation function of $p_{N,t}$ as
\begin{equation}\label{eq:corre-pt} p_{N,t}^{(k)} (x_1, \dots ,x_k) = \int d\by \, q_t^{(k)} (x_1, \dots , x_k; \by) \, p_{N,0} (\by) \end{equation}
where \begin{equation}\label{eq:kernel-pt} \begin{split} q_t^{(k)} (x_1, \dots , x_k ; \by) & =  \int q_t (\bx ; \by) \, d x_{k+1} \dots d x_N \\ & =\frac{(N-k)!}{N!} \; \det \left( K_{t,N} (x_i , x_j ; \by) \right)_{1 \leq i,j \leq k} \end{split} \end{equation}
with
\[
\begin{split}
  &K_{t,N} (u,v;\by) = \frac{N}{(2\pi i)^2 (v-u)t}
 \\ & \times  \int_\gamma dz \int_\Gamma dw \, (e^{-N(v-u)(w-r)/t} - 1) \prod_{j=1}^N 
  \frac{w-y_j}{z-y_j} \\
 &  \times \frac{1}{w-r}\Big( w-r+z-u - \frac{t}{N}\sum_j \frac{y_j-r}{(w-y_j)(z-y_j)}\Big)
  e^{N(w^2-2vw -z^2+2uz)/2t} \, .
\end{split}
\]
Here $\gamma$ is the union of two horizontal lines $\bR \ni s \to s-i\delta$, $\bR \ni s \to -s + i\delta$, for some $\delta >0$, and $\Gamma$ is a vertical line $\bR \ni s \to \kappa + is$ and $r \in \bR$ is arbitrary.

{F}rom (\ref{eq:corre-pt}), we observe that universality for matrices of the form (\ref{eq:Joh}) can be proven by showing that
\begin{equation}\label{eq:goal-Joh} \frac{1}{N\rho_{t} (E)} \, K_{t,N} \left(E + \frac{x_1}{N\rho_{t} (E)} , E + \frac{x_2}{N \rho_{t} (E)} ; \by \right)  \to \frac{ \sin \pi (x_2 - x_1)}{\pi (x_2 -x_1)} \end{equation}
for all $\by$ in a subset of $\bR^N$ whose $p_{N,0}$-measure tends to one, as $N \to \infty$. In (\ref{eq:goal-Joh}), \[ \rho_{t} (E) =\frac{1}{2\pi (1+t)^2} \sqrt{1- \frac{E^2}{4(1+t)^2}} \] is the rescaled semicircle law (the rescaling is needed because the variance of the entries of $H$ grows with $t$). 

To show (\ref{eq:goal-Joh}), Johansson writes 
\begin{equation}
\label{eq:kernel-as}
\begin{split}
 \frac{1}{N\rho_t (E)} K_{t,N} \Big(E,E+ &\frac{\tau}{N\rho_t (E)}; \by\Big) \\ &= N \int_\gamma \frac{d z}{2\pi i}\int_\Gamma \frac{d w}{2\pi i} \, 
 h_N(w) g_N(z,w) e^{N(f_N(w)-f_N(z))} \end{split} \end{equation}
with
\[ \begin{split}
   f_N(z) &= \frac{1}{2t}(z^2-2u z) +\frac{1}{N}\sum_j\log(z-y_j)
\\
  g_N(z,w) &= \frac{1}{t (w-r)}[w-r+z-u] -
\frac{1}{N(w-r)}\sum_j \frac{y_j-r}{(w-y_j)(z-y_j)}
\\
   h_N(w)  & =
\frac{1}{\tau} \Big( e^{-\tau (w-r)/t\varrho} - 1 \Big) \end{split}
\]
and he performs a detailed asymptotic analysis of the integral (\ref{eq:kernel-as}).
He finds the saddle points of the exponent $f_N (w) - f_N (z)$, he shifts the contours $\gamma, \Gamma$ to go through the saddles, he shows that the contributions of the saddles gives exactly the sine-kernel, and that the contributions away from the saddles vanish as $N \to \infty$. Following this strategy, Johansson proved convergence to the sine-kernel for ensembles of the form (\ref{eq:Joh}), for arbitrary fixed $t >0$. 

To prove universality for general ensembles of hermitian Wigner matrices, we would like to take $t=0$ in (\ref{eq:Joh}). As an intermediate step, we may ask what happens if one chooses $t = t(N)$ depending on $N$, so that $t(N) \to 0$ as $N \to\infty$. The algebraic identities discussed above are still valid, and universality follows again by showing that the r.h.s. of (\ref{eq:kernel-as}) converges to the sine-kernel as $N \to \infty$. The difference is that now the parameter $t$ entering the definition of $f_N(z), g_N (z,w), h_N (w)$ depends on $N$ and vanishes as $N \to \infty$; this makes the asymptotic analysis more delicate. It turns out, however, that, using Theorem \ref{thm:sc}, it is still possible to perform the asymptotic analysis of the integral in (\ref{eq:kernel-as}) and to show convergence to the sine-kernel, as long as $t(N) \geq N^{-1+\delta}$ for some $\delta >0$.Theorem \ref{thm:sc} is really the crucial ingredient of this analysis; the requirement $t(N) \geq N^{-1+\delta}$ is, in fact, a consequence of the fact that we only have convergence to the semicircle law on intervals of size $K/N$ or larger. The next theorem was proven in \cite{EPRSY}.
\begin{theorem}\label{thm:EPRSY1}
Let $H_0$ be an ensemble of hermitian Wigner matrices as in Def. \ref{def}. Choose a sequence $t(N)$ such that $t(N) \to 0$ as $N \to \infty$ and $t(N) \geq N^{-1+\delta}$ for some $\delta >0$. Let $V$ be a GUE matrix, independent of $H_0$. Then, the local eigenvalue statistics of $H = H_0 + t (N)^{1/2} V$ are such that
\begin{equation}\label{eq:WDtN} \frac{1}{\rho_{\text{sc}}^k (E)} p_N^{(k)} \left( E+ \frac{x_1}{N\rho_{\text{sc}} (E)}, \dots , E + \frac{x_k}{N\rho_{\text{sc}} (E)} \right) \to \det \left( \frac{\sin (\pi (x_i - x_j))}{(\pi (x_i -x_j))} \right)_{1 \leq i,j \leq k} \,  \end{equation}
for all $|E| < 2$. The convergence here holds after integrating against a bounded and compactly supported observable in the variables $x_1, \dots , x_k$.
\end{theorem}

%One may hope now that the Gaussian perturbation in $H = H_0 + t^{1/2} (N) V$ is so %small that it can not affect the local eigenvalue statistics. This would imply that also the %local correlations of the ensemble $H_0$ satisfy (\ref{eq:WDtN}).

One may hope to show that the small Gaussian perturbation cannot change the local eigenvalue correlations by comparing the laws of $H$ and $H_0$. If the entries of $H_0$ have the probability density function $h$, then the probability density function $h_t$ of the entries of $H_0 + t^{1/2} V$ is given by \[ h_t = e^{t L} h, \qquad \text{with } L = \frac{d^2}{dx^2}\,. \] At least formally, this implies that $|h_t -h| \simeq t Lh$. Hence, for small $t(N) \simeq N^{-1+\delta}$, the laws of the entries of $H$ and $H_0$ are very close. However, to compare the eigenvalue correlations of $H$ and of $H_0$, we would need to compare $N^2$ entries;  therefore, the total distance between the laws of $H$ and of $H_0$ is not small in the limit $N \to \infty$. To overcome this problem, we use a time-reversal type idea.  

Suppose that $H$ is an ensemble of hermitian Wigner matrices, whose entries have the probability density function $h$. 
%If possible, we would like to find a matrix $H_0$ whose entries have the probability %density function $h_0 = e^{-tL} h$; this would imply that $H = H_0 + t^{1/2} V$ and %universality for $H$ would follow from Johansson's argument. Of course, this is in %general not possible, because the heat flow cannot be inverted. Nevertheless, it is %possible to approximately invert the heat flow. For a given Wigner matrix $H$ with %entries having the probability density $h$, 
We can then introduce a new probability density function 
\begin{equation}\label{wth} \wt{h} = \left( 1-tL + \frac{t^2 L^2}{2} - \dots \pm \frac{t^n L^n}{n!} \right) h \end{equation}
and we can define a new Wigner matrix $\wt{H}$ whose entries have the law $\wt{h}$. The probability density function of the entries of $H_t = \wt{H} + t^{1/2} V$ is then given by
\[ h_t = e^{tL} \wt{h} = e^{tL}  \left( 1-tL + \frac{t^2 L^2}{2} - \dots \pm \frac{t^n L^n}{n!} \right) h\,. \]
Universality for the matrix $H_t$ follows from Theorem \ref{thm:EPRSY1} if $t = t(N) \geq N^{-1+\delta}$. Moreover, the density $h_t$ is now much closer to the initial $h$ compared with $e^{tL} h$. Formally, we find
\begin{equation}\label{eq:ht-h} |h_t - h| \simeq \frac{t^{n+1}}{(n+1)!} L^{n+1} h \, .\end{equation}
Taking $t = t(N) \simeq N^{-1+\delta}$, the r.h.s. can be made smaller than any power of $N$ by choosing $n \in \bN$ sufficiently large. Hence, if we choose $n$ large enough, we can compare the correlations of $H$ with the ones of $H_t$, and conclude universality for $H$ (from universality for $H_t$). {F}rom (\ref{eq:ht-h}), it is clear that, to choose $n$ large, we need to assume sufficient regularity of $h$; this explains the origin of the  regularity conditions in the next theorem, which appeared in \cite{EPRSY}.
%So, it is natural to ask how large do we have to choose $n$ to compare the correlations %of $H$ and $H_t$? It tunrs out that, in order to compare the $k$-point correlation %function of $H$ with the $k$-point correlation function of $H_t$, and therefore, in order %to show that the $k$-point correlation function of $H$ converges to the Wigner-Dyson %distribution, we need to take $n > k/2$. This observation leads to the following theorem, 
%which appeared in \cite{EPRSY}.
\begin{theorem}\label{thm:EPRSY2}
Let $H$ be an ensemble of hermitian Wigner matrices, whose entries have the probability density function $h (x) = e^{-g (x)}$. Fix $k \geq 1$ and assume $g \in C^{2(n+1)} (\bR)$ for some integer $n > k/2$, with 
\[ \sum_{j=1}^{2(n+1)} |g^{(j)} (x)| \leq C (1+ x^2)^m \] for some $m \in \bN$. Then we have 
\[ \frac{1}{\rho_{\text{sc}}^k (E)} p_N^{(k)} \left( E+ \frac{x_1}{N\rho_{\text{sc}} (E)}, \dots , E + \frac{x_k}{N\rho_{\text{sc}} (E)} \right) \to \det \left( \frac{\sin (\pi (x_i - x_j))}{(\pi (x_i -x_j))} \right)_{1 \leq i,j \leq k} \,  \]
for all $|E| < 2$ (after integrating against a bounded and compactly function $O(x_1, \dots , x_k)$).
\end{theorem}

%The conditions on the probability density function guarantees on the one hand the %regularity needed to expand, in (\ref{eq:wth}), up to the $n$-th order. On the other %hand, they guarantee that $h$ is everywhere positive and not too close to zero; if $h$ %is very close to zero, the function $h_t$ may loose its positivity, failing therefore to be a %probability density. 

Shortly after this result was posted, Tao and Vu obtained, in \cite{TV}, another proof of universality for hermitian ensembles of Wigner matrices. Their proof uses different techniques, but is also based on the convergence to the semicircle law on microscopic scales (more precisely, it is based on the delocalization of the eigenvectors of Wigner matrices, established in Theorem \ref{thm:deloc}). The result of Tao and Vu requires the third moment of the matrix entries to vanish, but, otherwise, almost no regularity.  Comparing the two works, we realized that the two approaches could be combined to yield an even stronger result. The following theorem was proven in \cite{ERSTVY}.

\begin{theorem}\label{thm:ERSTVY}
Let $H$ be an ensemble of Wigner matrices. Then, for every $k \in \bN$, $\delta >0$, $E_0 \in (-2 + \delta ; 2 - \delta)$, and for every compactly supported and bounded $O (x_1, \dots , x_k)$, we have 
\begin{equation}\label{eq:ERSTVY1}
 \begin{split} &\lim_{N \to \infty} \int_{E_0 - \delta}^{E_0+\delta} dE \int dx_1 \dots dx_k \; O(x_1, \dots , x_k) \\   &\times \left[ \frac{1}{\rho_{\text{sc}}^k (E)} p_N^{(k)} \left( E+ \frac{x_1}{N\rho_{\text{sc}} (E)}, \dots , E + \frac{x_k}{N\rho_{\text{sc}} (E)} \right) - \det \left( \frac{\sin (\pi (x_i - x_j))}{(\pi (x_i -x_j))} \right) \right] = 0 \, . \end{split} \end{equation}
If, moreover, $\E \, x_{ij}^3 = 0$ ($x_ij$ is the real part of the $(i,j)$-entry of $H$), we do not need to average over $E$. In other words, we have, for every fixed $|E| <2$, 
\begin{equation}\label{eq:ERSTVY2} \begin{split} \lim_{N \to \infty} &\int dx_1 \dots dx_k \, O(x_1, \dots  , x_k) \, \frac{1}{\rho_{\text{sc}}^k (E)} p_N^{(k)} \left( E+ \frac{x_1}{N\rho_{\text{sc}} (E)}, \dots , E + \frac{x_k}{N\rho_{\text{sc}} (E)} \right) \\ & = \int dx_1 \dots dx_k \; O (x_1, \dots , x_k) \, \det \left( \frac{\sin (\pi (x_i - x_j))}{(\pi (x_i -x_j))} \right)_{1 \leq i,j \leq k} \, . \end{split} \end{equation}
\end{theorem}

It is interesting to note that the results on universality we just discussed do not generalized easily to ensemble of Wigner matrices with different symmetry (real symmetric and quaternion hermitian ensembles). The reason is that the expressions (\ref{eq:qt-her}) and (\ref{eq:kernel-pt}) are based on a formula due to Harish-Chandra to integrate over the unitary group and, unfortunately, there is no analogous formula for integrating over the orthogonal (or the quaternion unitary) group. Also the result obtained in \cite{TV} by Tao and Vu can only be applied to ensembles of real symmetric or quaternion hermitian Wigner matrices if the first four moments of the matrix entries match exactly the corresponding Gaussian moments. To establish universality for real symmetric or quaternion hermitian ensembles of Wigner matrices, we developed new techniques, based on the introduction of a local relaxation flow, which approximates Dyson Brownian Motion, but is characterized by a faster relaxation time. In this case, we only prove universality after integrating the variable $E$ over an (arbitrarily small) interval (as in (\ref{eq:ERSTVY1})). Details can be found in \cite{ESY4}.

\section{Extensions and an Application of Universality}
\label{sec:ados}

In \cite{P}, P{\'e}ch{\'e} shows universality (pointwise in $E$) for  ensembles of complex {\it sample covariance matrices} extending the approach outlined in Section \ref{sec:univ}. In \cite{ESYY}, we prove universality for ensembles of (real or complex) sample covariance matrices (but only after integration over $E$ in a small interval). 

In \cite{EYY1,EYY2}, Erd\H os, Yau and Yin extend the results presented above to so called {\it generalized Wigner matrices}. Up to the symmetry constraints, the entries of generalized Wigner matrices are independent, but not necessarily identically distributed. Instead, one assumes that the variances $\sigma_{ij} = \E \, x_{ij}^2$, $1 \leq i,j \leq N$ are such that $c_1 \leq N \sigma_{ij} \leq c_2$. 

In \cite{EYY3}, the same authors show that the eigenvalues of generalized Wigner matrices are localized with very high probability within distances of order $(\log N)^{\alpha}/N$ from the position where the semicircle law predicts they should be. 

More related results are also presented in the review paper \cite{E}.

To conclude these notes, we mention a new result concerning the {\it average density of states} for hermitian Wigner matrices. Theorem \ref{thm:sc} establishes the convergence of the density of states to the semicircle law on the microscopic scale. On shorter scale we cannot expect the density of states to converge
in probability, because its fluctuations are certainly important. 
%; in fact, consider the interval $I_{\eps} = [E-(\eps/2N) ; E+ (\eps / 2N)]$, for a %sufficiently small $\eps >0$. The density of states in $I_{\eps}$ is typically equal to zero %(because, typically, the interval $I_{\eps}$ will contain no eigenvalue). With small %probability (of order $\eps$, in fact), the interval $I_{\eps}$ will contain one or more %eigenvalues, and the density of states will be very large, of order $1/\eps$. In any %event, the fluctuations of the density of states with respect to the semicircle law are at %least of order one, and therefore convergence in probability cannot hold true. What can %we say, however, if we take the expectation of the density of states? 
Nevertheless, we may still ask whether on these extremely short scales the expectation of the density of states, known as the average density of states, converges. The next theorem (see \cite{MS}) gives a positive answer to this question.  
\begin{theorem}\label{thm:ados}
Let $H$ be an ensemble of hermitian Wigner matrices. Assume that the probability density of the matrix entries has the form $h (x) = e^{-g(x)}$ and satisfies the bounds 
\[ \left| \widehat{h} (p) \right| \leq \frac{1}{(1 + C p^2)^{\sigma/2}}\, ,\quad \left| \widehat{h g''} (p) \right| \leq \frac{1}{(1 + C p^2)^{\sigma/2}} \quad \text{for some } \sigma > 6 .\]
Then, for any $|E| <2$, we have 
\begin{equation}\label{eq:ados0} \frac{1}{\eps} \, \E \, \cN \, \left[ E-\frac{\eps}{2N} ; E+ \frac{\eps}{2N} \right] \to \rho_{\text{sc}} (E) \end{equation}
as $N \to \infty$, {\it uniformly} in $\eps >0$. In other words, we have
\[ \lim_{N \to \infty} \limsup_{\eps \to 0} \, \E \, \frac{ \cN \, \left[ E-\frac{\eps}{2} ; E+ \frac{\eps}{2} \right]}{N\eps} = \lim_{N \to \infty} \liminf_{\eps \to 0} \, \E \, \frac{ \cN \, \left[ E-\frac{\eps}{2} ; E+ \frac{\eps}{2} \right]}{N\eps} =  \rho_{\text{sc}} (E). \]
%In particular, this means that, for an arbitrary sequence $\delta (N) >0$ such that $
%\delta (N) \to 0$ as $N \to \infty$, 
%\begin{equation} \label{eq:ados} \lim_{N \to \infty} \E \, \frac{\cN \left[E-\frac{\delta(N)}
%{2} ; E+ \frac{\delta(N)}{2} \right]}{N \delta (N)} = \rho_{\text{sc}} (E) . \end{equation}
\end{theorem}
Universality (in the form (\ref{eq:ERSTVY2})) implies that 
\begin{equation}\label{eq:uni-sc} \E \, \frac{\cN\left[E-\frac{\kappa}{2N} ; E + \frac{\kappa}{2N} \right]}{\kappa} = \int \frac{{\bf 1} (|x| \leq \kappa /2)}{\kappa} \, p^{(1)}_{N} (E + \frac{x}{N}) \to \rho_{sc} (E) \, . \end{equation} for an arbitrary fixed ($N$ independent) $\kappa >0$. Hence, to prove Theorem \ref{thm:ados}, we need to understand how (\ref{eq:uni-sc}) can be extended to intervals of size $\eps /N$ with $\eps$ going to zero as $N \to \infty$ (or even $\eps \to 0$ before $N \to \infty$). Eq. (\ref{eq:ados0}) follows, if we can show 
\begin{equation}\label{eq:EmN} \frac{1}{\pi}  \, \E \; \text{Im } m_N \left(E+i \frac{\eps}{N} \right) \to \rho_{\text{sc}} (E) \end{equation} uniformly in $\eps >0$. To this end, we establish an upper bound of the form 
\begin{equation}\label{eq:dEmN} \left| \frac{d}{d E} \, \E \; \text{Im } m_N \left(E+i \frac{\eps}{N} \right) \right| \leq C N \end{equation} 
uniformly in $\eps >0$. {F}rom (\ref{eq:dEmN}), it follows that $\E \, \text{Im } m_N (E+i(\eps/N))$ remains essentially constant if $E$ varies within an interval of size $\kappa/N$, for a small (but fixed) $\kappa >0$. This means that 
\[ \begin{split}
\frac{1}{\pi} \, \E \; \text{Im } & m_N \left( E+i \frac{\eps}{N} \right) \\ &\simeq  \E  \, \frac{N}{\pi \kappa}  \, \int_{E- \frac{\kappa}{2N}}^{E+\frac{\kappa}{2N}}  d E' \,   \text{Im }  m_N \left(E' + i \frac{\eps}{N} \right) 
\\ &= \E \, \frac{1}{\pi \kappa} \sum_{\alpha} \left[ \text{arctg} \,  \left( \frac{N \left(\mu_\alpha - E - \frac{\kappa}{2N}\right)}{\eps} \right) - \text{arctg} \, \left( \frac{N \left(\mu_\alpha - E + \frac{\kappa}{2N}\right)}{\eps} \right) \right] \\ &\simeq \E \, \frac{1}{\kappa} \, \cN \left[E-\frac{\kappa}{2N} ; E+ \frac{\kappa}{2N} \right] \to \rho_{sc} (E) \, 
\end{split} \]
as $N \to \infty$, by (\ref{eq:uni-sc}). It remains to show the upper bound (\ref{eq:dEmN}); here we use a Wegner estimate from \cite{ESY3}, which implies that the average density of states remains uniformly bounded on arbitrarily small intervals. The same techniques used in \cite{ESY3} to show the Wegner estimate can be extended to obtain the bound (\ref{eq:dEmN}). Details can be found in \cite{MS}.


\thebibliography{hhhhh}


\bibitem{BP} Ben Arous, G., P\'ech\'e, S.: Universality of local
eigenvalue statistics for some sample covariance matrices.
{\it Comm. Pure Appl. Math.} {\bf LVIII.} (2005), 1--42.

\bibitem{D}  Dyson, F.J.: Correlations between eigenvalues of a random
matrix. {\it Commun. Math. Phys.} {\bf 19}, 235-250 (1970).

\bibitem{E} Erd\H os, L.: Universality of Wigner random matrices: a Survey of Recent Results. Preprint arXiv:1004.0861.

\bibitem{EPRSY}
Erd\H{o}s, L.,  P\'ech\'e, S.,  Ram\'irez, J.,  Schlein,  B.,
and Yau, H.-T., Bulk universality 
for Wigner matrices.
{\it Commun. Pure Appl. Math.} {\bf 63}, No. 7,  895--925 (2010)


\bibitem{ERSTVY}
Erd\H{o}s, L., Ram\'irez, J., Schlein, B., Tao, T., Vu, V. and Yau, H.-T.:
Bulk universality for Wigner hermitian matrices with subexponential decay.
{\it Int. Math. Res. Notices.} {\bf 2010}, No. 3, 436-479 (2010)


\bibitem{ERSY}  Erd{\H o}s, L., Ramirez, J., Schlein, B., Yau, H.-T.:
 Universality of sine-kernel for Wigner matrices with a small Gaussian
perturbation.
{\it Electr. J. Prob.} {\bf 15},  Paper 18, 526--604 (2010)


\bibitem{ESY1} Erd{\H o}s, L., Schlein, B., Yau, H.-T.:
Semicircle law on short scales and delocalization
of eigenvectors for Wigner random matrices.
{\it Ann. Probab.} {\bf 37}, No. 3, 815--852 (2009)


\bibitem{ESY2} Erd{\H o}s, L., Schlein, B., Yau, H.-T.:
Local semicircle law  and complete delocalization
for Wigner random matrices. {\it Commun.
Math. Phys.} {\bf 287}, 641--655 (2009)

\bibitem{ESY3} Erd{\H o}s, L., Schlein, B., Yau, H.-T.:
Wegner estimate and level repulsion for Wigner random matrices.
{\it Int. Math. Res. Notices.} {\bf 2010}, No. 3, 436-479 (2010)

\bibitem{ESY4} Erd{\H o}s, L., Schlein, B., Yau, H.-T.: Universality
of random matrices and local relaxation flow. Preprint. arXiv:0907.5605

\bibitem{ESYY} Erd{\H o}s, L., Schlein, B., Yau, H.-T., Yin, J.:
The local relaxation flow approach to universality of the local statistics for random matrices. Preprint arXiv:0911.3687.

\bibitem{EYY1} Erd{\H o}s, L.,  Yau, H.-T., Yin, J.: Bulk
 universality for generalized Wigner matrices. 
Preprint. 	arXiv:1001.3453.

\bibitem{EYY2} Erd{\H o}s, L.,  Yau, H.-T., Yin, J.: 
Universality for generalized Wigner matrices with Bernoulli distribution.
Preprint arXiv:1003.3813.

\bibitem{EYY3} Erd{\H o}s, L.,  Yau, H.-T., Yin, J.: 
Rigidity of Eigenvalues of Generalized Wigner Matrices.
Preprint arXiv:1007.4652.

\bibitem{J} Johansson, K.: Universality of the local spacing
distribution in certain ensembles of Hermitian Wigner matrices.
{\it Comm. Math. Phys.} {\bf 215} (2001), no.3. 683--705.

\bibitem{MS} Maltsev, A., Schlein, B., paper in preparation.

\bibitem{P}
P\'ech\'e, S., Universality in the bulk of the spectrum for complex sample covariance matrices. Preprint, arXiv:0912.2493.

\bibitem{S} Soshnikov, A.: Universality at the edge of the spectrum in
Wigner random matrices. {\it  Comm. Math. Phys.} {\bf 207} (1999), no.3.
697-733.

\bibitem{TV} Tao, T. and Vu, V.: Random matrices: Universality of the
local eigenvalue statistics. Preprint arXiv:0906.0510.

\bibitem{TV2} Tao, T. and Vu, V.: Random matrices: 
Random matrices: Universality of local eigenvalue statistics up to the edge.
Preprint arXiv:0908.1982.

\bibitem{TW} Tracy, C. and Widom, H.: Level-spacing distribution and Airy kernel. 
{\it Commun. Math. Phys.} {\bf 159} (1994), 151Ð174.

\bibitem{W} Wigner, E.: Characteristic vectors of bordered matrices
with infinite dimensions. {\it Ann. of Math.} {\bf 62} (1955), 548-564.
\end{document}